\documentclass[times]{aa}
\usepackage{epsfig}
\def\Msun{\hbox{$\rm\thinspace M_{\odot}$}}
\newcommand{\lapprox }{{\lower0.8ex\hbox{$\buildrel <\over\sim$}}}
\def\la{\mathrel{\hbox{\rlap{\hbox{\lower4pt\hbox{$\sim$}}}{\raise2pt\hbox{$<$}}
}}}
\def\ga{\mathrel{\hbox{\rlap{\hbox{\lower4pt\hbox{$\sim$}}}{\raise2pt\hbox{$>$}}
}}}

\begin{document}

   \thesaurus{06(08.19.4 SN 1006)}
                           
%
   \title{The Schweizer$-$Middleditch star revisited}

   \author{M. R. Burleigh \inst{1} \and U. Heber \inst{2} 
\and D. O'Donoghue \inst{3} \and M. A. Barstow \inst{1}}

   \offprints{Matt Burleigh, mbu@star.le.ac.uk}

   \institute{Department of Physics and Astronomy, University of
                Leicester, Leicester LE1 7RH, UK
\and  Dr. Remeis-Sternwarte Bamberg, Universit\"at
Erlangen-N\"urnberg, Sternwartstrasse 7, D-96049 Bamberg, Germany
\and South African Astronomical Observatory, PO Box 9, Observatory 7935,
Cape Town, South Africa }


   \date{Received 30 November 1999 /  Accepted 9 February 2000}

    \titlerunning{The Schweizer$-$Middleditch Star Revisited} 

    \authorrunning{M. R. Burleigh et~al.}

\maketitle

\begin{abstract}

We have re-observed and re-analysed the optical spectrum of the 
Schweizer$-$Middleditch star, a hot subdwarf which lies along almost 
the same line-of-sight as the centre of the historic SN1006 supernova
remnant (SNR). 
Although this object is itself unlikely to be the remnant of the star 
which exploded in 1006AD, Wellstein et~al. (1999) have demonstrated that
it could be the remnant of the donor star in a pre-supernova Type Ia 
interacting binary, if it possesses an unusually low mass. We show that,
if it had a mass of 0.1$-$0.2$\Msun$, 
the SM star would lie at the same distance
($\approx$800pc) as the SNR as estimated by Willingale et~al.
(1995). However, most distance estimates to SN1006 are much larger than
this, and there are other convincing arguments to 
suggest that the SM star lies behind this SNR. Assuming the 
canonical subdwarf mass of 0.5$\Msun$, we constrain the distance of the 
SM star as 1050~pc$<$d$<$2100~pc. This places the upper limit on the 
distance of SN1006 at 2.1~kpc.

   \end{abstract}

%

\section{Introduction}

SN 1006 was the brightest supernova witnessed in recorded history. The
estimated peak magnitude (V$=$$-$9.5$\pm$1, Clark \& Stephenson
1977), reported visibility for nearly two years, and the lack of a nearby
OB association strongly suggests a Type Ia origin (SNIa, Minkowski 1966).
Almost all current models of Type Ia supernova involve the nuclear
explosion of a white dwarf induced by rapid mass accretion in a binary
system. However, no stellar remnant from this supernova explosion 
has ever been conclusively identified, 
including a pulsar, or the remains of any companion star.

\begin{figure*}
\vspace{8.5cm}
\includegraphics{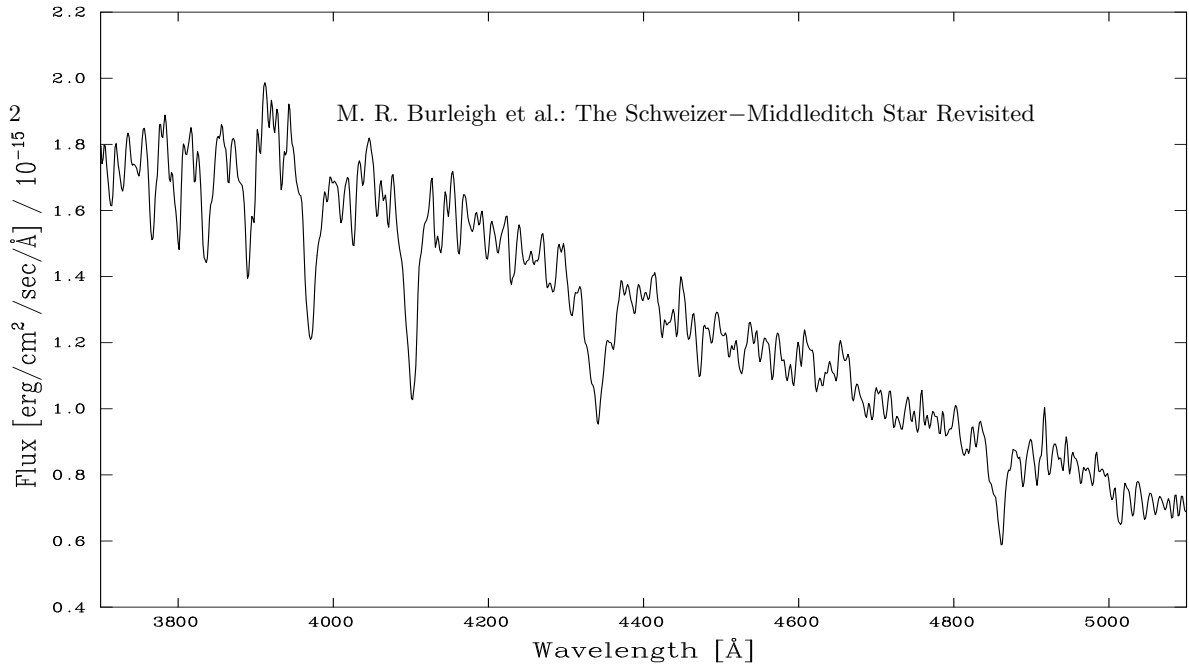}
\caption{Optical spectrum of the SM star, smoothed with a 3 pixel Gaussian.}
\end{figure*}

In 1980, Schweizer \& Middleditch searched for just such 
a stellar remnant from SN 1006  and discovered a faint (V$=$16.7) 
blue star $\approx$2.5' from the projected 
centre of the supernova remnant (SNR). 
They identified this object (now known as the
Schweizer$-$Middleditch star, SM star or SM80) 
as a hot subdwarf sdOB
star, and estimated its effective temperature 
T$_{eff}$$=$38,500$\pm$4500K, and surface 
gravity log g$=$6.7$\pm$0.6. From an estimate
of the absolute magnitude, M$_v$$=$6.2$\pm$1.8,  Schweizer and
Middleditch (1980) derived a  distance to their subdwarf of 1.1 ($+$1.4,
$-$0.6) kpc. 
Since chance projection seemed unlikely, and the distance estimate was in
rough agreement with the then exisiting estimates of the distance to the
SNR itself, Schweizer \& Middleditch (1980) suggested that
their subdwarf may in fact 
be the remnant star, or at least associated with it. 

Savedoff and Van Horn (1982) later 
showed conclusively that the SM star could not be the remnant of the 
supernova itself, since the time to cool to the observed effective 
temperature was simply too long, $\sim$10$^6$ years compared to the SNR age of 
10$^3$ years. However, this does not rule out the SM star as a stellar remnant 
of the {\it donor} star in a pre-SNIa interacting binary system.

Subsequent far ultra-violet (far-UV) 
observations with IUE and HST/FOS revealed the presence
of strong Fe II and Si II, III and IV lines 
superimposed on the continuum of the SM star (Wu et~al. 1983, Fesen
et~al. 1988, Wu et~al. 1993). The iron lines have symmetrical 
velocity profiles, broadened up to $\sim$8000 km
s$^{-1}$ FWHM. The Si features are asymmetric, redshifted  
and centred at a radial
velocity of $\sim$5000 km s$^{-1}$. These features have been used to
estimate the mass of iron in the remnant and to map the positions of
various shock regions. Importantly, though, the presence of redshifted
lines in the supernova ejecta suggests that the SM star must
lie {\it behind} the SNR, since they are assumed to originate in material 
moving away from us on the far side of the remnant.

Measurements of the widths of these aborption lines, coupled with the
angular size of the remnant, led Wu et~al. (1993) to derive a {\it lower
limit} to the SNR distance of 1.9 kpc. This contrasts strongly with the
estimate of Willingale et~al. (1995) of 0.7$\pm$0.1 kpc, derived from 
modelling X-ray emission detected in 
ROSAT PSPC observations. Therefore, we were motivated to
re-observe and re-analyse the SM star in order to place tighter
constraints on its distance, and hence on the distance to the SNR itself.
Secondly, we learnt of the study by Wellstein et~al. (1999) 
which suggests that the prior donor star in an
SN Ia progenitor system (an interacting binary) may appear subsequently
as a {\it low mass} hot subdwarf star. This new theoretical result re-opens 
the question first posed by Schweizer \& Middleditch (1980) in the conclusion 
to their discovery paper: "Can one component of a
binary system that forms a Type Ia supernova end up being a hot subdwarf
or white dwarf?". In the light of Wellstein et~al.'s recent 
work, we re-address this question.

\section{Spectroscopy}

The SM star was observed for a total of 
4000 seconds on 1996 April 14 with the South
African Astronomical Observatory's 1.9-m Ratcliffe Telescope, the Unit
spectrograph and the Reticon photon counting system (RPCS). The RPCS had
two arrays, one which accumulates energy from the source, while the other
records sky background through an adjacent aperture. The target was
observed for 2000 seconds through one aperture, then for a further 2000
seconds through the second aperture, in order to average out variations
between the two light paths. The grating (number
6) was blazed to cover a wavelength range of 
$\sim$3700{\AA}$-$5200{\AA} with a
resolution of $\approx$4{\AA}. Flat fields were obtained at the start and
end of the night, and wavelength calibration was provided by a CuAr lamp, 
which was observed before and after the target. A blue spectro-photometric
standard (LTT 6248) was also observed. The reduced, calibrated spectrum is 
shown in Fig. 1.

\section{High speed photometry}

\begin{figure}
\vspace{11cm}
\includegraphics{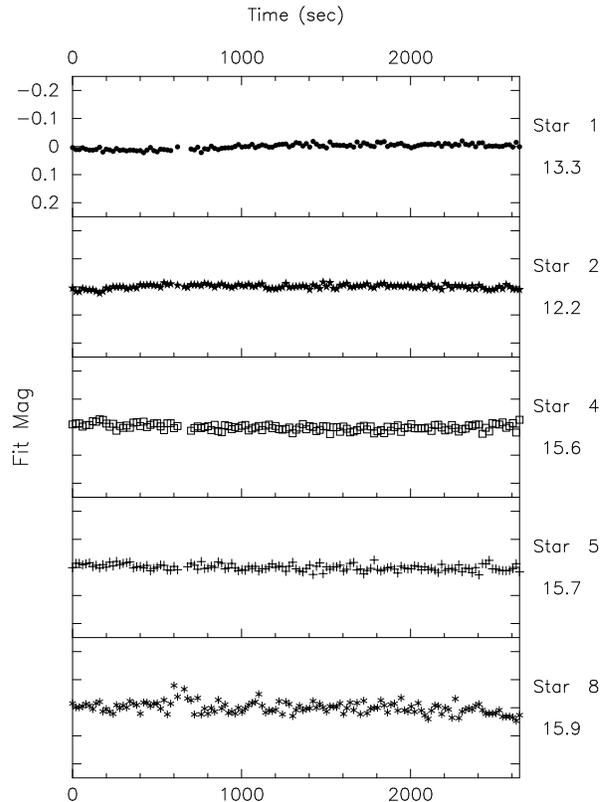}
\caption{Differential light curve for the SM Star (\#8) and four 
comparison stars in the same field.}
\end{figure}

\begin{figure}
\vspace{6cm}
\includegraphics{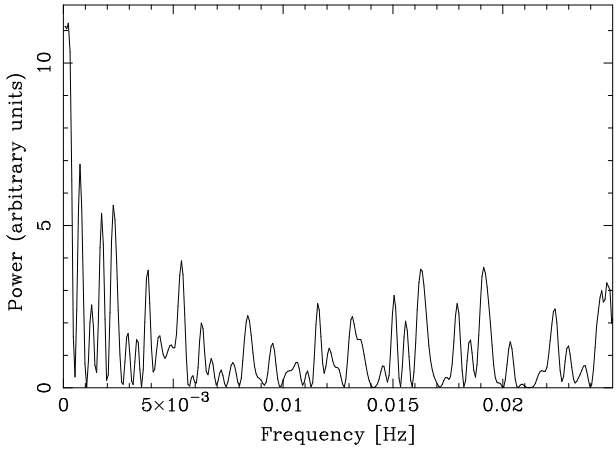}
\caption{Amplitude spectrum determined from the SM Star's light curve.}
\end{figure}

Recently, multi-periodic pulsations have been discovered in a number of 
subdwarf sdB stars (the EC14026 stars, Kilkenny et~al., 1997). Both radial 
and non-radial modes are present, although the cause of these pulsations 
is not fully understood. 
Theoretical studies have shown that these oscillations may be excited by 
an opacity bump due to heavy element ionization, giving rise to a 
metal-enrichment in this driving region (Charpinet et~al., 1996). 
However, why pulsations 
are observed in some sdBs and not in others remains a mystery. 

We observed the SM Star on 1999 September 4th with the South African 
Astronomical Observatory's 0.75m telescope, together with the 
University of Cape Town's CCD photometer in high speed mode, in 
order to search for pulsations. A $\approx$2600 second light curve 
was obtained, consisting of 20 second exposures separated 
by essentially zero seconds of dead time. 
Four comparison stars were also observed at the same time. 
The differential light curve is shown in Fig. 2. 
The SM star (star \#8 in Fig. 2) shows no evidence of 
pulsations; the fluctuations in Fig. 2 are merely random noise. 
The amplitude spectrum (Fig. 3), which has been calculated out to the 
Nyquist frequency, also shows no evidence for pulsations. However, 
at V$\approx$16.7 we are clearly unable to detect fluctuations below 
$\approx$0.05 mags. with this telescope. Many of the known sdB pulsators 
vary at the level of 0.001$-$0.05 mags., and so clearly we cannot rule out 
low level pulsations in this object. We suggest that it should be re-observed 
on a larger telescope.

\section{Analysis}

\subsection{Spectral analysis}

The H Balmer series is
visible in the calibrated optical spectrum (Fig. 1)
to H11. HeI is detected at 4026{\AA}, 4144{\AA}, 4472{\AA} and
marginally at 4922{\AA}. There is also a marginal detection of HeII at
4686{\AA}. 

A grid of synthetic spectra derived from H \& He line blanketed NLTE 
model atmospheres (Napiwotzki 1997) was matched to the data to 
simultaneously determine the effective temperature, surface gravity 
and He abundance (see Heber et~al. 1999). We find  T$_{eff}$$=$32,900K, 
log g$=$6.18 and log (N(He)$/$N(H))$=$$-$1.7. While formal statistical errors 
from the fitting procedure are relatively small  (1$\sigma$:
$\Delta$(T$_{eff}$$=$)=340K, $\Delta$(log\,g)$=$$\Delta$(log(He/H))$=$0.1dex),
systematics dominate the error budget and are estimated from varying the 
spectral windows for the profile fitting and the continuum setting to be 
$\Delta$(T$\_{eff}$)$=$$\pm$1500K, $\Delta$(log g)$=$$\pm$0.3 dex and 
$\Delta$(log (N(He)$/$N(H))$=$$\pm$0.3 dex. These best-fit 
parameters are unchanged if H$_\epsilon$ is omitted 
from the fit (since it might be contaminated by CaII). 
A more precise error estimate 
would, however, require repeat observations.

Therefore, we find that both the temperature and 
gravity are at the low end of the large range estimated 
by Schweizer \& Middleditch (1980). With these parameters the
SM star resembles an ordinary subdwarf B star 
close to the zero-age extended horizontal branch (ZAEHB).

\subsection{Extinction}

Using the Matthews \& Sandage (1963) calibration, combined with our model 
fit parameters, we estimate the colour excess E$_{(B-V)}$$=$0.16$\pm$0.02. 
From Whitford (1958) we then estimate the visual extinction 
A$_v$$=$3.0$\times$E$_{(B-V)}$$=$0.48$\pm$0.06. Schweizer \& Middleditch 
measured the V magnitude from photoelectric photometry as 16.74$\pm$0.02. 
Therefore, we take the 
redening corrected magnitude as 
\newline V$_0$$=$16.26$\pm$0.07.

\subsection{Distance}

Since bolometric corrections for hot subluminous stars are large and 
somewhat uncertain, we prefer not to make use of them for the distance 
determination. Instead we calculate the angular radius from the 
ratio of the observed (dereddened) flux at the effective wavelength of the 
V filter and the corresponding model flux.
Assuming the canonical mass for hot subdwarf 
stars, M$=$0.5$\Msun$, we determine the stellar radius from the 
gravity and finally derive the distance from the angular diameter and the 
stellar radius. 
We obtain a distance of d$=$1485pc which corresponds to an absolute 
magnitude of M$_V$$=$5.4. However, the error on log g 
is large ($\pm$0.3 dex), translating to d$=$1050pc for log g$=$6.48, or 
d$=$2100pc for log g$=$5.88. 

If the SM star has a much lower mass than usually assumed for these 
objects, as suggested by Wellstein 
et~al. (1999), then the absolute magnitude will be lower 
and hence the star will be much closer to us. For example, 
if M$=$0.2$\Msun$ then we find M$_V$$=$6.4 and 
d$=$940pc (assuming log g$=$6.18).  
If M$=$0.1$\Msun$, then M$_V$$=$7.2 and d$=$650pc.

\begin{figure}
\vspace{11cm}
\includegraphics{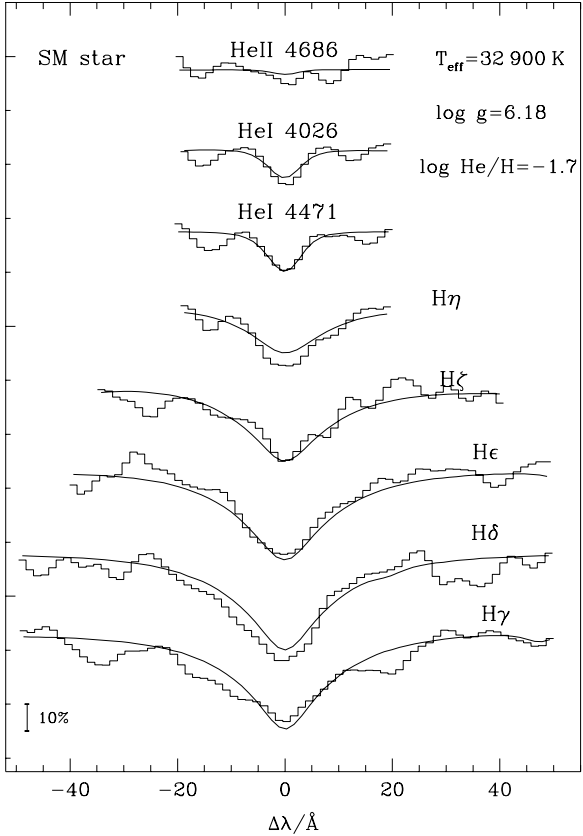}
\caption{NLTE model fit to the H Balmer lines and He lines detected in the 
SM Star's optical spectrum.}
\end{figure}

\begin{figure}
\vspace{11cm}
\includegraphics{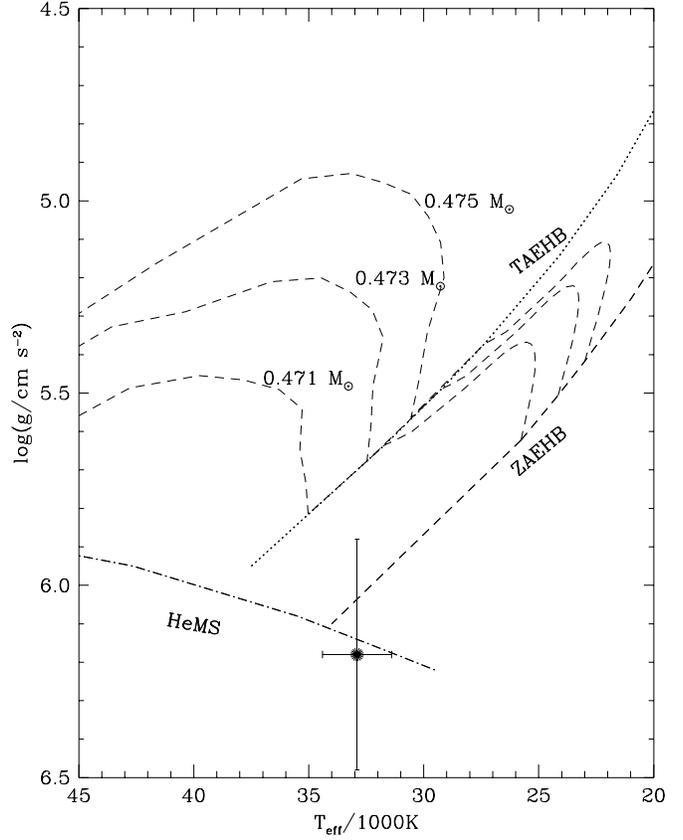}
\caption{Position of the SM star in the T$_{eff}$/log g plane (large cross). 
The zero-age 
extended horizontal branch (ZAEHB) 
and He main sequence are marked. Loci showing how 
stars of various masses evolve away from the ZAEHB are also shown.}

\end{figure}

\section{Discussion}

A new analysis of the Schweizer-Middleditch star, a hot subdwarf which lies 
along the same line-of-sight as the centre of the SN1006 SNR, has allowed 
us to place tighter constraints on its atmospheric parameters, and re-assess 
its distance. Since Wellstein et~al. (1999) have demonstrated that the 
remnant of 
the donor star in a pre-SNIa binary system could appear as a hot subdwarf, 
albeit with an abnormally low mass, we can now re-address Schweizer \&
Middleditch's original question: is the SM star the stellar remnant of one 
component of the SNIa progenitor binary?

In order to begin answering this question, we need to convice ourselves
that the SM star lies at the same distance as the SN1006 SNR.
Unfortunately, there is a large range in the SNR distance estimates
quoted in the literature. 
In Table 1, we list the various distance estimates to the SN1006 SNR 
itself and the method used to obtain that distance. Early estimates, 
based for example on the historical record of its brightness (e.g. Minkowski 
1966) and early models of the X-ray emission, gave distances $\sim$1kpc. 
Most of the more recent estimates, based on a variety of theoretical
models or measurements of e.g.~the expansion velocity or proper motion of
optical filaments, place the SNR at a distance of $\sim$1.5$-$2.0kpc. The
one glaring exception is the estimate of Willigale et~al. (1995),
0.7$\pm$0.1kpc, based on an analysis of the ROSAT PSPC X-ray image of the
SNR. 

We find the distance to the SM star 1050$<$d$<$2100 pc, 
assuming that it is an
ordinary hot subdwarf. If Willingale et~al's distance estimate is correct,
then 
the SM star would lie a long way behind the remnant. In order
for it to lie within the remnant, it would have to be of unusually low
mass. A mass of 0.1$-$0.2$\Msun$ gives a distance compatible with 
Willingale et~al's estimate, and  
in that scenario the SM star could indeed then be
a remnant of the donor star in an SNIa progenitor system. 

However, if Willingale et~al's SNR  distance estimate is wildly inaccurate, 
and 
the more conservative estimates of $\sim$1.5$-$2.0kpc are correct, then
the SM star cannot be a low mass remnant of the donor star in a 
pre-SNIa binary.

\begin{table*}
\begin{center}
\caption{Distance estimates to the SN1006 SNR from the literature}
\begin{tabular}{lcll}
Author & Year & Distance & Method \\
       &      &   (kpc)  &        \\
Minkowski & 1966 & 1.3 & Historical record of brightness \\
Stephenson et~al. & 1977 & 1.0$\pm$0.3 & Historical record of brightness \\
Winkler & 1977 & 0.9$-$1.3 & Reverse shock model of x-ray emission \\
Hamilton et~al. & 1986 & 1.7 & Reverse shock model   \\
Kirshner et~al. & 1987 & 1.4$-$2.1 & Shock velocity \& proper motions \\
Hamilton \& Fesen & 1988 & 1.5$-$2.0 & Spherically symmetric hydrodynamic 
simulations \\
Fesen et~al. & 1988 & 1.5$-$2.3 & Fe line widths, age \& angular size of 
remnant \\
Long et~al. & 1988 & 1.7$-$3.1 & Proper motion of optical filaments \\
Wu et~al. & 1993 & $>$1.9 & FeII line widths \& angular size of 
remnant \\
Willingale et~al. & 1995 & 0.7$\pm$0.1 & Analysis of x-ray emitting material \\
Laming et~al. & 1996 & 1.8$\pm$0.3 & Modelling non-radiative shocks \\
\end {tabular}
\end{center}
\end{table*}

In fact, there are two more compelling arguments against the SM star
having any relation to SN1006. Firstly, it is located $\approx$2.5' south
of the projected centre of the remnant, and would have to possess a
proper motion of 0.15" per year and a velocity of $\approx$800km
sec$^{-1}$ to have reached its current location. Unfortunately, the star
simply does not possess this motion or velocity. Secondly, the presence
of red-shifted metal absorption lines superimposed on the SM star's UV 
spectrum strongly indicate that the star is behind the remnant, since
these features almost certainly originate at a shock front on the
remnant's far side. Confirmation of this may come 
from observations of other nearby objects with strong UV fluxes and generally
featureless far-UV continuums. Indeed, P.F. Winkler has an HST/STIS
program to observe four such objects behind SN1006 during Cycle 8
(two QSOs and two A0 stars, program ID 8244), and one of these objects
is even closer to the projected centre of SN1006 than the SM star. 
These targets are not scheduled to be observed until
June-July 2000, but the detection of the same red-shifted features 
as seen in the SM star (and the non-detection of any 
additional features with separate velocities) would effectively
rule out any exotic origin for these lines, and confirm the location of
the SM star behind the SN1006 SNR.

Thus, the SM star can only be the remnant of the donor star in a pre-SNIa
binary, such as might have produced SN1006, if the following four
criteria are fulfilled: (1) The star has an unusually low mass for a hot
subdwarf ($\approx$0.1$\Msun$), (2) the low distance estimate to the SN1006
SNR of Willingale et~al. (1995) is correct, (3) the red-shifted metal
lines seen in the SM star's far-UV spectrum originate somewhere other
than on the far side of the SNR, and (4) the SM star has a high proper
motion and transverse velocity. Unfortunately, at the time of writing,
none of these conditions can convincingly shown to be true.
However, the tighter constraint we have been able to place on the distance 
to the
SM star in this analysis can now be used to place an upper limit on the
distance to the SN1006 SNR itself, and hence constrain the models and
methods used to estimate the distances of supernova remnants.

\begin{acknowledgements}

MRB acknowledges the support of PPARC, UK. We thank Pete Wheatley (Leicester 
University) for generating the Fourier transform of the SM Star's light 
curve.

\end{acknowledgements}

\end{document}